\documentclass[superscriptaddress,prl,twocolumn,amsmath,amssymb,floatfix,showpacs]{revtex4-1}
\usepackage{graphicx}
\usepackage{bm}
\usepackage{amssymb}
\usepackage{color}
\usepackage{epsfig}
\usepackage{subfigure}

\usepackage{amsmath}
\usepackage{hyperref}
\usepackage{bbold}
\hypersetup{
    pdfstartview={FitH},
   pdfpagemode={None}
   }
\newcommand{\be}{\begin{equation}}
\newcommand{\ee}{\end{equation}}
\begin{document}



\title{Thermal conductivity in 1d: disorder-induced transition from anomalous to normal scaling}

\author{Ariel Amir}
\affiliation{School of Engineering and Applied Sciences, Harvard University, Cambridge, Massachusetts 02138, USA}
\author{Yuval Oreg}
\author{Yoseph Imry}
\affiliation {Department of Condensed Matter Physics, Weizmann Institute of Science, Rehovot, 76100, Israel}

\begin{abstract}

It is well known that the contribution of harmonic phonons to the thermal conductivity of 1D systems diverges with the harmonic chain length $L$ (explicitly, increases with $L$ as a power-law with a positive power). Furthermore, within various one-dimensional models containing disorder it was shown that this divergence persists, with the thermal conductivity scaling as $\sqrt{L}$ under certain boundary conditions, where $L$ is the length of the harmonic chain. Here we show that when the chain is weakly coupled to the heat reservoirs and there is strong disorder this scaling can be violated. We find a weaker power-law dependence on $L$, and show that for sufficiently strong disorder the thermal conductivity stops being anomalous -- despite both density-of-states and the diverging localization length scaling anomalously. Surprisingly, in this strong disorder regime two anomalously scaling quantities cancel each other to recover Fourier's law of heat transport.
\end{abstract}


 \maketitle

Over the past decades, studying the thermal conductance of low-dimensional systems has generated much interest both theoretically and experimentally, and both at the classical and quantum level \cite{rubin1971abnormal, casher1971heat, rego1998quantized, dhar2001heat, chen2005nanoscale,chaudhuri2010heat,dhar2015heat, luckyanova2016phonon}. Intriguingly, the thermal conductivity of an ordered lattice diverges, in any dimension. This comes about since the density-of-states (DOS) of an ordered systems scales as $\omega^{d-1}$ at low frequencies (where $d$ is the dimensionality of the system), while the Rayleigh scattering length scales as $\frac{1}{\omega^{d+1}}$. It can be shown that the contribution of a narrow frequency range to the thermal conductivity scales as the product of these two quantities, and the thermal conductivity is given by an integral over all frequencies. The low-frequency divergence of the integral implies an infinite conductivity at any dimension. For a weakly disordered system in one-dimension, it can be shown that for a finite system size $L$ the thermal conductivity scales as $\sqrt{L}$ \cite{casher1971heat,dhar2001heat, dhar2015heat}(under certain boundary conditions, to be elaborated on later on). This raises the question: can strong disorder ``fix" the anomalous dependence of thermal conductivity on system size? Here we show that for disorder possessing a ``heavy-tail" (with diverging moments) the answer is affirmative.

\emph{The model.-} Consider a chain of harmonic springs, constructed as follows. First, $N$ points are chosen randomly and uniformly in the interval [0,L], and a mass $M$ is placed on each of them. Without loss of generality we choose $L=N-1$, such that the average nearest-neighbor distance $r_{nn}=1$. Next, the springs between nearest-neighbor masses are chosen as:
\be K = e^{-r/\xi} , \ee
with $r$ the distance and $\xi$ a constant. The dimensionless parameter $\epsilon \equiv \xi/r_{nn}$ is a measure of the disorder, where for $\epsilon \to 0$ the distribution of spring constants becomes very broad, corresponding to the case of strong disorder. Since the nearest-neighbor distance follows a Poisson process, we find that:
\be P(K) \propto K^{\epsilon-1}. \label{spring_dist} \ee
The spectrum and localization properties of this model were studied in Refs. \cite{alexander1981excitation, ziman, Exp_Mat}, and are reviewed below.

Consider coupling the system now to two infinite, thermal baths, at temperatures $\mathbf{T_1}$ and $\mathbf{T_2}$. Within each of the bath, phonons of all frequencies would be propagating in both directions. The partitioning of phonon energies is given by the Bose-Einstein distribution, and each of them is transmitted through the disordered region with some probability $T(\omega)$ (depending on the properties of the disordered region as well as its coupling to the baths).
Accounting for the density-of-states in the leads, the heat flux going from left to right is found to be \cite{rego1998quantized}:

\be \dot Q = \frac{1}{2\pi} \int_0^\infty T(\omega) N_{\mathbf{T_1}}(\omega) \hbar \omega  d\omega, \ee
where $T(\omega)$ is the transmission probability of a phonon with frequency $\omega$ and $N_{\mathbf{T}}$ is the Bose-Einstein function.
By writing a similar expression for the right-to-left heat flux and taking the limit $\mathbf{T_1-T_2} \to 0$, we find that the heat conductance $G$ (distinct from the conductivity, $\sigma$) is:

\be G =  \frac{1}{2\pi}\int_0^\infty T(\omega) \frac{\partial N_{\bar{\mathbf{T}}}(\omega)}{\partial \bar{\mathbf{T}}}\Big|_{\mathbf{T}} \hbar \omega  d\omega.\ee

Note that here $T(\omega)$ is the frequency-dependent transmission, while ${\mathbf{T}}$ is the temperature.
Later we will see that in the thermodynamic limit $L \to \infty$ the dominant contribution to this integral comes from low frequency phonons, in which case we can approximate $\frac{\partial N_{\bar{\textbf{T}}}(\omega)}{\partial \bar{\textbf{T}}} \approx \frac{k_B}{\hbar \omega}$. Therefore we find the conductance is given by:

\be G =  \frac{k_B}{2\pi}\int_0^\infty T(\omega) d\omega . \label{sigma_int} \ee

A similar result is obtained in Ref. \cite{dhar2001heat}.

\begin{figure}[hbt!]
\centering\includegraphics[width=0.5\textwidth]{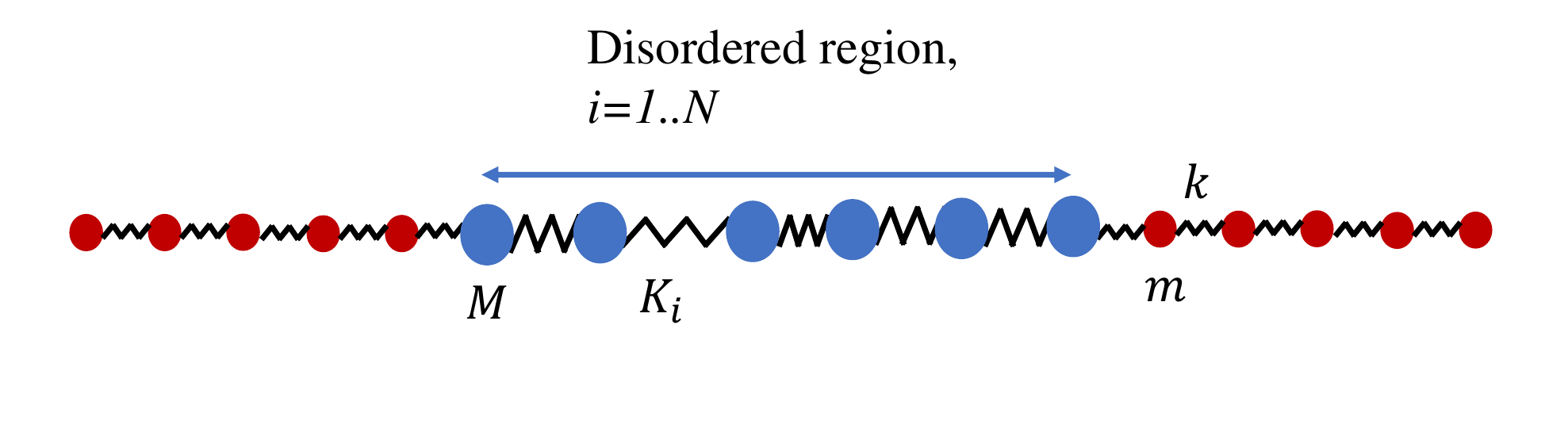}
\caption{Schematic illustration of the model. The red circles correspond to the continuum (masses $m$, spring constant $k$), while masses with $i=2...N-1$ correspond to the disordered region, where the springs are drawn from a power-law distribution (Eq. (\ref{spring_dist}), while the masses are constant and equal $M$. }\label{model_fig}
\end{figure}

Consider now a model in which the bath consists of an infinite set of identical masses $m$ and springs $k$, with lattice constant $a$. Masses in the disordered region will have indices of $1$ to $N$. Masses $1$ and $N$ will be connected to the bath via a spring of strength $k$. The model is illustrated schematically in Fig. \ref{model_fig}. We shall now show that the transmission through this chain is well approximated by a sum of narrow Lorentzians, with different widths and areas. Thus, they contribute non-uniformly to the integral of Eq. (\ref{sigma_int}).

The transmission can be readily expressed in terms of the disordered spring constants. Consider a wave incoming from the LHS. Away from the disordered region, the amplitude scales as $a(x) = e^{i q x} + r e^{-i q x}$, where for convenience we set $x=0$ for the $0$th mass, while to the right of the disordered region we have $t e^{i q x}$, where here we choose the amplitude of the first mass in the ordered region (with index $N+1$) to be $t$. For the masses in the disordered region, denoted by indices $1..N$, the equations of motion read:

\be M\omega^2 x_i  + K_{i}[x_{i+1}-x_i] + K_{i-1}[x_{i-1}-x_i] =0, \ee

where as stated above $K_0=K_N=k$.



We can recast this set of equations in a more concise matrix form:

\be [A+M \omega^2 I] \vec{x} = \vec{b}, \ee

where the matrix $A$ is tri-diagonal with elements: $A_{ii} = -K_{i-1} - K_i$, $A_{i,i-1}=K_{i-1}$, $A_{i,i+1}=K_i$, and all components of the vector $\vec{b}$ vanish save for $\vec{b}_1 = -k x_0 $ and $\vec{b}_N =-k x_{N+1}$, with $k$ the spring constant in the ordered region.

According to our assumptions we have: $x_{N+1}=t$, $x_{0}=1+r$. Finally, to close the set of equations we write the equations-of-motion for masses $0$ and $N+1$:

\begin{align}
 m \omega^2 x_0  + k [x_{1}-x_0] + k[x_{-1}-x_0] &=0 ; \\
   m \omega^2 x_{N+1}  + k [x_{N}-x_{N+1}] + k[x_{N+2}-x_{N+1}] &=0. \nonumber
\end{align}

Finally, we have $x_{N+2}= e^{i \chi} t$, $x_{-1} = e^{-i \chi} + r e^{i \chi}$,

where $\chi= (\omega/c)a$, $a$ being the spacing between masses in the ordered chain and $c$ the speed of sound.

To proceed, notice that we can express $\vec{x}$ in terms of the eigenmodes $ |v_\lambda \rangle$ and eigenvalues $\lambda$ of $A$, as:

\be  \vec{x} = [A+ M \omega^2 I]^{-1} \vec{b} = \sum_\lambda  \frac{1}{\lambda+M\omega^2} \langle v_\lambda | \vec{b} \rangle |v_\lambda \rangle ,\label {sum_trans}\ee
where we are using the quantum-mechanical notation for convenience, and assumed that the eigenmodes are normalized. In the following we shall also assume that the eigenmodes entries are real, which we can assume without loss of generality since $A$ is Hermitian.

If the springs in the ordered region are sufficiently small, the transmission will be negligible for nearly every frequency, except when $\omega$ is close to an eigenfrequency of the chain (in which case the denominator vanishes). We will assume this to be the case, and later on we will establish the precise condition on $k$ and $m$ for this to hold (note that this is one particular way of realizing the thermal bath). When the driving frequency is close to this resonant frequency by a detuning $\delta \lambda = M (\omega^2 - \omega_0^2)$, only this particular eigenmode $|v \rangle$ will contribute to the sum of Eq. (\ref{sum_trans}), leading to:

\be \vec{x} \approx \frac{1}{\delta \lambda} \langle v | \vec{b} \rangle |v \rangle. \ee

%
%
%

These equations determine $r$ and $t$. Solving leads to:


\be t =\frac{-k v_1 v_n (1- e^{-2 i \chi}]}{e^{2i\chi}[e^{-i\chi}{\delta \lambda}+k({|v_N|^2 }+{|v_1|^2 })]}.   \label{res_t}\ee

Note that, as expected, in order for the transmission to be non-negligible the eigenmode should have support both at the beginning and end of the disordered chain. i.e., only eigenmodes which have localization lengths comparable or larger to the length of the chain contribute to the transmission, and hence to the thermal conductance. Later we shall show that for $L \to \infty$, only the low-lying modes will be delocalized, hence we can assume the frequency $\omega$ and thus $\chi$ to be small. Hence: 

\be T=|t|^2 \approx  \frac{4k^2 \chi^2 (v_1 v_N)^2}{\chi^2 {\delta \lambda}^2+(\delta \lambda+k [v_1^2+v_N^2] )^2} ,\ee

leading to:

\be T \approx \frac{4k^2 \chi^2(v_1 v_N)^2}{{(\delta \lambda+S)}^2+\chi^2 S^2} ,\label{trans}\ee
where $S\approx k [v_1^2+v_N^2]$ and $\delta \lambda = -M (\omega^2 - \omega_0^2)$. It is useful to replace $(\delta \lambda+S)$ with $-M(\omega^2-\tilde{\omega}^2) \approx -2 M \tilde{\omega} (\omega-\tilde{\omega})$, which shows that this form is approximately a \emph{Lorentzian} in terms of $\omega$:

\be  T(\omega) \approx \frac{4k^2 \chi^2(v_1 v_N)^2}{{[2 M \tilde{\omega} (\omega-\tilde{\omega})]}^2+\chi^2 S^2} ,\label{trans2}\ee

\begin{figure}[hbt!]
\centering\includegraphics[width=0.5\textwidth]{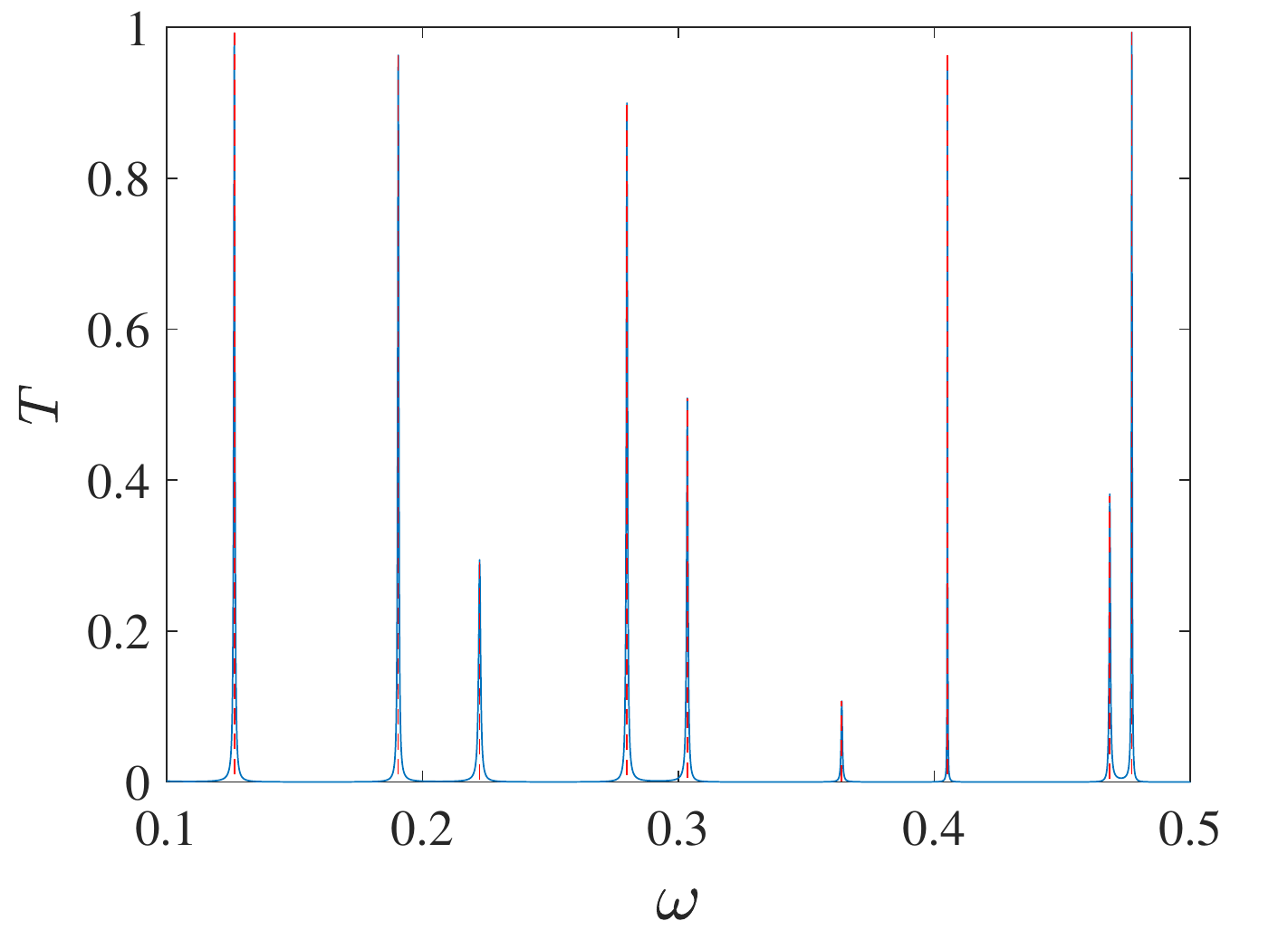}
\caption{Numerical simulation of a weakly disordered chain of length $N=50$. The red vertical lines show the eigenfrequencies of the disordered region, which nearly coincide with the Lorentzian peaks in the transmission. The area under each Lorentzian is determined by the product of the amplitudes of the right and left masses, and increases with the degree of ``delocalization" of the mode. The red dashed lines indicate the eigenfrequencies of the disordered chain, and their height corresponds to the numerator of Eq. (\ref{res_t}).}\label{example_modes}
\end{figure}

An example of this is shown in Fig. \ref{example_modes}. The maximal transmission is governed by the asymmetry of the eigenmode, and is 1 for $v_1=v_N$. The area associated with each Lorentzian is:

\be \Sigma_\lambda = 2 \pi k_B \frac{\sqrt{k m}}{M }\frac{(v_1 v_N)^2}{v_1^2+v_N^2}.\ee
Therefore it depends on the product $\frac{(v_1 v_N)^2}{v_1^2+v_N^2}$ but not explicitly on the frequency. Specifically, the contribution of all delocalized modes to the conductance is \emph{constant}. For a purely delocalized mode (on the scale of the disordered region, $L$) we have $v_1=v_N= \frac{1}{\sqrt{N}}$ and thus the contribution to the thermal conductance is $\Sigma_\lambda = \frac{2 \pi k_B \sqrt{k m} }{M N}$.
Note that this constant is different than the ``quantum of thermal conductance" \cite{rego1998quantized}, and depends on the properties of the bath (through $k$ and $m$).
Next, we will consider how the disorder affects the scaling of the thermal conductance, by summing over the contribution of all modes which are effectively delocalized.

{Application to disordered chain.} Within the aforementioned model, in the thermodynamic limit phonons of all frequencies are localized, with a diverging length scale at $\omega \to 0$. As noted above, for a finite system of size $L$, according to Eq. (\ref{trans}) only phonons will localization length of order or larger than $L$ will contribute to the thermal conductance, with each mode contributing a constant amount independent of frequency. Thus we can write:

\be G \approx \int_0^{\omega_c} \nu(\omega) d\omega , \label{sigma_final}\ee

where $\nu(\omega)$ is the density-of-states (DOS) of the phonons in the \emph{disordered region} and the localization length at $\omega_c$ equals $L$. (Note that the DOS is proportional to $N$).

According to Ref. \cite{ziman}, for disorder below a critical strength ($\epsilon \geq 1$ in our above definitions) we have \emph{Debye} DOS (constant in 1d) and for $\epsilon \leq 2$ the localization length diverges as \cite{ziman2}:

\be l_{loc}(\omega) \propto  1/\omega^2 . \label {weak_dis} \ee

This result -- corresponding to Rayleigh scattering in one dimension -- has been also derived in Refs. \cite{connor, baluni} for low-lying acoustic modes with weak disorder of a different type. Moreover, since in one-dimensional systems the localization length equals the mean free path \cite{thouless1973localization}, this result reflects the Rayleigh-like nature of scattering in the weakly disordered regime. 

Plugging this into Eq. (\ref{sigma_final}) we find:

\be G (L) \propto 1/\sqrt{L} .\ee

This implies that the thermal \emph{conductivity} $\sigma$ diverges as $\sqrt{L}$, a result obtained in the context of several other related disordered models \cite{rubin1971abnormal, dhar2001heat, dhar2015heat}.

Intriguingly, this result changes when we consider strong disorder. For $1 \leq  \epsilon \leq 2$, the DOS follows a Debye spectrum, while the localization length diverges more slowly in this regime  \cite{ziman}:

\be l_{loc}(\omega) \propto  1/\omega^\epsilon . \label{loc_anomalous}\ee

Note that it is precisely at the point $\epsilon =2$ that the variance in the compressibility of the system becomes ill-defined: the effective spring constant is the sum of $1/K_i$, hence the compressibility is related to $\langle 1/K \rangle$. The distribution of $z=1/K$ follows $p(z)\propto 1/z^{1+\zeta}$, hence it becomes heavy-tailed at the point $\zeta=2$. Similarly, at the point $\zeta=1$ the mean of this distribution diverges making the compressibility ill-defined in the continuum limit -- as we shall now see, the behavior will again dramatically change at this point. This is reminiscent of the qualitative change in the diffusion properties in the context of anomalous diffusion \cite{klafter2011first} as well as the emergence of aging in Bouchaud's trap model when the mean trapping time diverges \cite{bouchaud1992weak}.

Eq. (\ref{loc_anomalous}) implies that:

\be G (L)  \propto 1/L^{1/\epsilon}.\ee

Thus, the thermal conductance becomes normal (i.e., obeying Fourier's law) as opposed to anomalous at the point $\epsilon=1$.

Finally, for even stronger disorder, $\epsilon \leq 1$, we have  \cite{ziman}:

\be \nu(\omega)  \propto \omega^{\frac{\epsilon-1}{\epsilon+1}} , \label{DOS_dis} \ee

Thus, for strong disorder the DOS develops a strong singularity at the origin approximately diverging as $1/\omega$. This ``boson-peak" like behavior as well as the localization length scaling in this regime were derived using a strong-disorder renormalization group procedure in Ref. \cite{Exp_Mat}. The localization length in this case scales as:

\be l_{loc}(\omega) \sim 1/\omega^{\frac{2\epsilon}{1+\epsilon}} .\ee

This implies that:

\be G (L) \propto 1/L .\ee

\begin{figure}[t!]
\centering\includegraphics[width=0.4\textwidth]{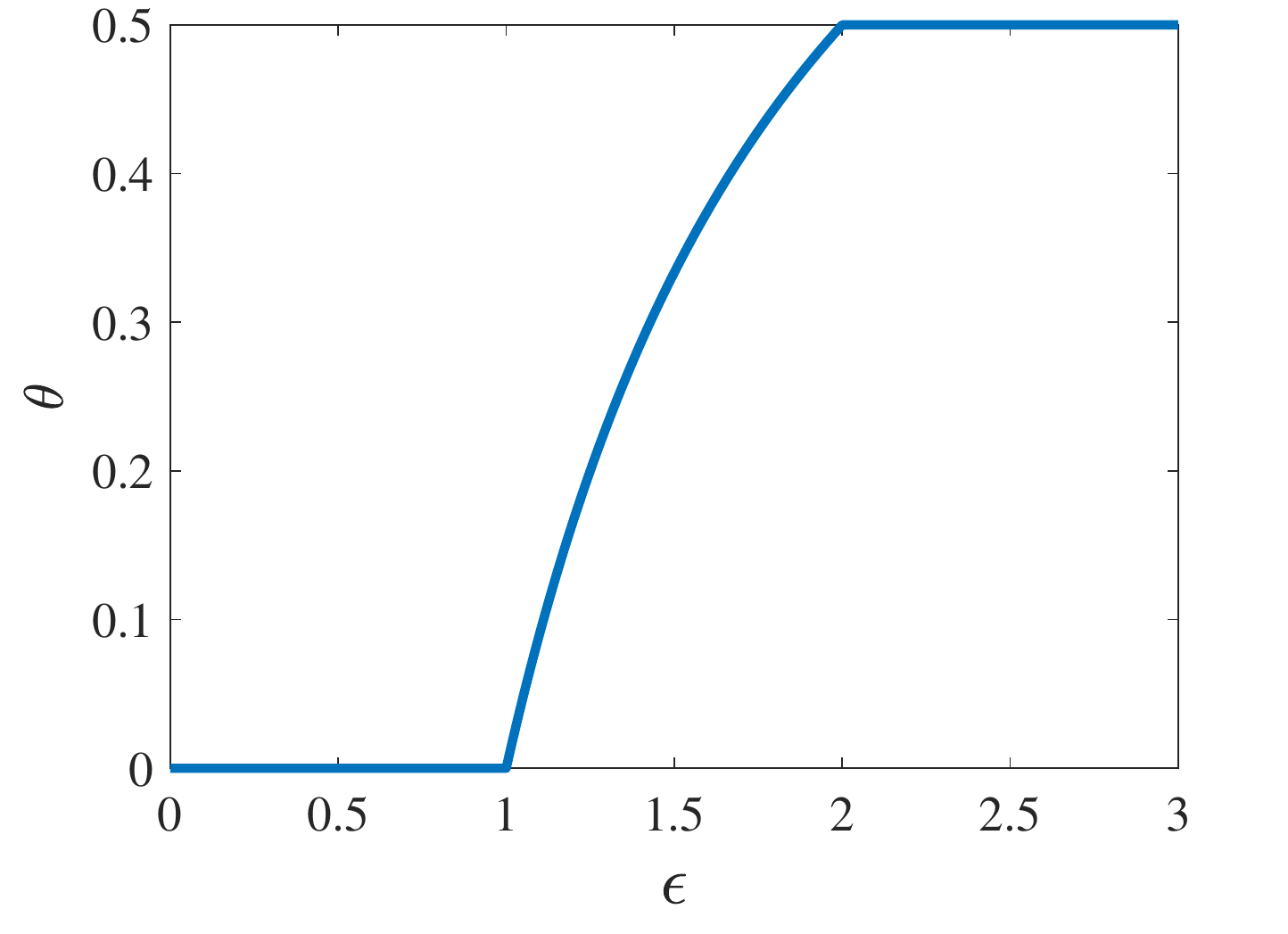}
\caption{Scaling of the conductivity $\sigma$ on system size $L$. We find that the thermal conductivity depends on the size of the system $L$ in a power law fashion $\sigma \propto L^\theta$. The exponent $\theta$ and its dependence on the disorder strength $\epsilon$ are depicted in the Figure. For weak disorder (large $\epsilon$), an anomalous scaling of $\sqrt{L}$ is observed, similar to results found in models with mass-disorder. Intrudingly, the model with spring disorder shows a phase-transition to a strong-disorder regime where the scaling of the thermal conductivity with system size is no longer anomalous. }\label{results}
\end{figure}

Thus, for disorder above a critical threshold the divergence of the thermal conductivity is remedied, and the thermal conductivity $\sigma$ becomes independent of system size. These results are summarized in Fig. \ref{results}, and in Table 1, which constitute our main results.

\begin{table*}[t!]
\caption{Summary of expected exponents for thermal conductance (from this work), DOS and localization length (from Ref. \cite{ziman}). } 
\centering 
\begin{tabular}{c c c c c c} 
\hline\hline 
Disorder strength  & DOS exponent $\propto \omega^\alpha$ & Localization length exponent $\propto 1/\omega^\beta$ & Thermal conductivity exponent $\propto L^\theta$\\ [0.5ex] 
\hline 
$\epsilon>2$ (weak disorder) & 0 (Debye) & 2 (Rayleigh) & 1/2 (anomalous) \\ 
$1 < \epsilon\leq 2$ (intermediate disorder) & 0 (Debye) & $\epsilon$ & $1-1/\epsilon$ (anomalous)\\
$\epsilon \le 1$ (strong disorder) & $\frac{\epsilon-1}{\epsilon+1}$ & $\frac{2\epsilon}{1+\epsilon}$ & 0 (normal)\\ [1ex] 
\hline 
\end{tabular}
\label{table:nonlin} 
\end{table*}

We may now revisit our previous assumption -- that a \emph{single} eigenmode contributes to each transmission peak. This is equivalent to demanding that the Lorentzians are well separated. The width of each of the delocalized modes can be readily identified from Eq. (\ref{trans2}) to be $\frac{\sqrt{k m}}{ N M}$ (for all disorder regimes). Their separation is the inverse of the local DOS. For the weak disorder regime, the DOS is constant and scales as $\frac{1}{N}\sqrt{\frac{K}{M}}$, where $K=1$ is the strongest possible spring within the model. Hence the condition for the applicability of the approximation is $k/K \ll M/m$.  The same is true in the intermediate disorder, since the DOS is still constant. However for the strong disorder regime $\epsilon < 1$ the DOS is given by Eq. (\ref{DOS_dis}), hence the condition becomes $\frac{\sqrt{k m}}{ N M} \ll  \omega^{\frac{1-\epsilon}{\epsilon+1}}\sqrt{\frac{K}{M}}$. This condition has to be fulfilled for the typical delocalized frequency $\omega_c \sim N^{-1/\epsilon}$, giving a more stringent (N-dependent) condition for the ratio $\frac{k m}{K M}$.

To summarize, we have studied a simple model of phonon conduction in 1D harmonic chain, and found intriguing behavior of the thermal conductivity. While for weak disorder we reproduce the known $\sqrt{L}$ scaling of the conductivity with system size, we found that for ``heavy-tailed" disorder this scaling changes, with a scaling exponent that changes smoothly until the disorder power-law reaches a critical threshold at which the Fourier law is recovered -- despite density-of-states and localization length both scaling differently than in the weak disorder regimes (i.e., they do not follow the Debye and Rayleigh laws). Further work will establish what physical system are described by a weak coupling to the thermal bath as studied here, and what happens in the case of stronger coupling, where interference between the different eigenmodes is significant, potentially via the use of numerical simulations. Furthermore, in the future it would be interesting to study how strong disorder affects the phonon thermal conductivity in higher dimensions, and what the fate of Fourier's law is in that scenario.

\emph{Acknowledgments} We thank Jie Lin, Yohai Bar-Sinai and Bertrand Halperin for useful discussions.


\begin{thebibliography}{17}%
\makeatletter
\providecommand \@ifxundefined [1]{%
 \@ifx{#1\undefined}
}%
\providecommand \@ifnum [1]{%
 \ifnum #1\expandafter \@firstoftwo
 \else \expandafter \@secondoftwo
 \fi
}%
\providecommand \@ifx [1]{%
 \ifx #1\expandafter \@firstoftwo
 \else \expandafter \@secondoftwo
 \fi
}%
\providecommand \natexlab [1]{#1}%
\providecommand \enquote  [1]{``#1''}%
\providecommand \bibnamefont  [1]{#1}%
\providecommand \bibfnamefont [1]{#1}%
\providecommand \citenamefont [1]{#1}%
\providecommand \href@noop [0]{\@secondoftwo}%
\providecommand \href [0]{\begingroup \@sanitize@url \@href}%
\providecommand \@href[1]{\@@startlink{#1}\@@href}%
\providecommand \@@href[1]{\endgroup#1\@@endlink}%
\providecommand \@sanitize@url [0]{\catcode `\\12\catcode `\$12\catcode
  `\&12\catcode `\#12\catcode `\^12\catcode `\_12\catcode `\%12\relax}%
\providecommand \@@startlink[1]{}%
\providecommand \@@endlink[0]{}%
\providecommand \url  [0]{\begingroup\@sanitize@url \@url }%
\providecommand \@url [1]{\endgroup\@href {#1}{\urlprefix }}%
\providecommand \urlprefix  [0]{URL }%
\providecommand \Eprint [0]{\href }%
\providecommand \doibase [0]{http://dx.doi.org/}%
\providecommand \selectlanguage [0]{\@gobble}%
\providecommand \bibinfo  [0]{\@secondoftwo}%
\providecommand \bibfield  [0]{\@secondoftwo}%
\providecommand \translation [1]{[#1]}%
\providecommand \BibitemOpen [0]{}%
\providecommand \bibitemStop [0]{}%
\providecommand \bibitemNoStop [0]{.\EOS\space}%
\providecommand \EOS [0]{\spacefactor3000\relax}%
\providecommand \BibitemShut  [1]{\csname bibitem#1\endcsname}%
\let\auto@bib@innerbib\@empty
\bibitem [{\citenamefont {Rubin}\ and\ \citenamefont
  {Greer}(1971)}]{rubin1971abnormal}%
  \BibitemOpen
  \bibfield  {author} {\bibinfo {author} {\bibfnamefont {R.~J.}\ \bibnamefont
  {Rubin}}\ and\ \bibinfo {author} {\bibfnamefont {W.~L.}\ \bibnamefont
  {Greer}},\ }\href@noop {} {\bibfield  {journal} {\bibinfo  {journal} {Journal
  of Mathematical Physics}\ }\textbf {\bibinfo {volume} {12}},\ \bibinfo
  {pages} {1686} (\bibinfo {year} {1971})}\BibitemShut {NoStop}%
\bibitem [{\citenamefont {Casher}\ and\ \citenamefont
  {Lebowitz}(1971)}]{casher1971heat}%
  \BibitemOpen
  \bibfield  {author} {\bibinfo {author} {\bibfnamefont {A.}~\bibnamefont
  {Casher}}\ and\ \bibinfo {author} {\bibfnamefont {J.}~\bibnamefont
  {Lebowitz}},\ }\href@noop {} {\bibfield  {journal} {\bibinfo  {journal}
  {Journal of Mathematical Physics}\ }\textbf {\bibinfo {volume} {12}},\
  \bibinfo {pages} {1701} (\bibinfo {year} {1971})}\BibitemShut {NoStop}%
\bibitem [{\citenamefont {Rego}\ and\ \citenamefont
  {Kirczenow}(1998)}]{rego1998quantized}%
  \BibitemOpen
  \bibfield  {author} {\bibinfo {author} {\bibfnamefont {L.~G.~C.}\
  \bibnamefont {Rego}}\ and\ \bibinfo {author} {\bibfnamefont {G.}~\bibnamefont
  {Kirczenow}},\ }\href {\doibase 10.1103/PhysRevLett.81.232} {\bibfield
  {journal} {\bibinfo  {journal} {Phys. Rev. Lett.}\ }\textbf {\bibinfo
  {volume} {81}},\ \bibinfo {pages} {232} (\bibinfo {year} {1998})}\BibitemShut
  {NoStop}%
\bibitem [{\citenamefont {Dhar}(2001)}]{dhar2001heat}%
  \BibitemOpen
  \bibfield  {author} {\bibinfo {author} {\bibfnamefont {A.}~\bibnamefont
  {Dhar}},\ }\href {\doibase 10.1103/PhysRevLett.86.5882} {\bibfield  {journal}
  {\bibinfo  {journal} {Phys. Rev. Lett.}\ }\textbf {\bibinfo {volume} {86}},\
  \bibinfo {pages} {5882} (\bibinfo {year} {2001})}\BibitemShut {NoStop}%
\bibitem [{\citenamefont {Chen}(2005)}]{chen2005nanoscale}%
  \BibitemOpen
  \bibfield  {author} {\bibinfo {author} {\bibfnamefont {G.}~\bibnamefont
  {Chen}},\ }\href@noop {} {\emph {\bibinfo {title} {Nanoscale energy transport
  and conversion: a parallel treatment of electrons, molecules, phonons, and
  photons}}}\ (\bibinfo  {publisher} {Oxford University Press},\ \bibinfo
  {year} {2005})\BibitemShut {NoStop}%
\bibitem [{\citenamefont {Chaudhuri}\ \emph {et~al.}(2010)\citenamefont
  {Chaudhuri}, \citenamefont {Kundu}, \citenamefont {Roy}, \citenamefont
  {Dhar}, \citenamefont {Lebowitz},\ and\ \citenamefont
  {Spohn}}]{chaudhuri2010heat}%
  \BibitemOpen
  \bibfield  {author} {\bibinfo {author} {\bibfnamefont {A.}~\bibnamefont
  {Chaudhuri}}, \bibinfo {author} {\bibfnamefont {A.}~\bibnamefont {Kundu}},
  \bibinfo {author} {\bibfnamefont {D.}~\bibnamefont {Roy}}, \bibinfo {author}
  {\bibfnamefont {A.}~\bibnamefont {Dhar}}, \bibinfo {author} {\bibfnamefont
  {J.~L.}\ \bibnamefont {Lebowitz}}, \ and\ \bibinfo {author} {\bibfnamefont
  {H.}~\bibnamefont {Spohn}},\ }\href@noop {} {\bibfield  {journal} {\bibinfo
  {journal} {Physical Review B}\ }\textbf {\bibinfo {volume} {81}},\ \bibinfo
  {pages} {064301} (\bibinfo {year} {2010})}\BibitemShut {NoStop}%
\bibitem [{\citenamefont {Dhar}\ and\ \citenamefont
  {Dandekar}(2015)}]{dhar2015heat}%
  \BibitemOpen
  \bibfield  {author} {\bibinfo {author} {\bibfnamefont {A.}~\bibnamefont
  {Dhar}}\ and\ \bibinfo {author} {\bibfnamefont {R.}~\bibnamefont
  {Dandekar}},\ }\href {\doibase http://dx.doi.org/10.1016/j.physa.2014.06.002}
  {\bibfield  {journal} {\bibinfo  {journal} {Physica A: Statistical Mechanics
  and its Applications}\ }\textbf {\bibinfo {volume} {418}},\ \bibinfo {pages}
  {49 } (\bibinfo {year} {2015})},\ \bibinfo {note} {proceedings of the 13th
  International Summer School on Fundamental Problems in Statistical
  Physics}\BibitemShut {NoStop}%
\bibitem [{\citenamefont {Luckyanova}\ \emph {et~al.}(2016)\citenamefont
  {Luckyanova}, \citenamefont {Mendoza}, \citenamefont {Lu}, \citenamefont
  {Huang}, \citenamefont {Zhou}, \citenamefont {Li}, \citenamefont {Kirby},
  \citenamefont {Grutter}, \citenamefont {Puretzky}, \citenamefont
  {Dresselhaus} \emph {et~al.}}]{luckyanova2016phonon}%
  \BibitemOpen
  \bibfield  {author} {\bibinfo {author} {\bibfnamefont {M.~N.}\ \bibnamefont
  {Luckyanova}}, \bibinfo {author} {\bibfnamefont {J.}~\bibnamefont {Mendoza}},
  \bibinfo {author} {\bibfnamefont {H.}~\bibnamefont {Lu}}, \bibinfo {author}
  {\bibfnamefont {S.}~\bibnamefont {Huang}}, \bibinfo {author} {\bibfnamefont
  {J.}~\bibnamefont {Zhou}}, \bibinfo {author} {\bibfnamefont {M.}~\bibnamefont
  {Li}}, \bibinfo {author} {\bibfnamefont {B.~J.}\ \bibnamefont {Kirby}},
  \bibinfo {author} {\bibfnamefont {A.~J.}\ \bibnamefont {Grutter}}, \bibinfo
  {author} {\bibfnamefont {A.~A.}\ \bibnamefont {Puretzky}}, \bibinfo {author}
  {\bibfnamefont {M.~S.}\ \bibnamefont {Dresselhaus}},  \emph {et~al.},\
  }\href@noop {} {\bibfield  {journal} {\bibinfo  {journal} {arXiv preprint
  arXiv:1602.05057}\ } (\bibinfo {year} {2016})}\BibitemShut {NoStop}%
\bibitem [{\citenamefont {Alexander}\ \emph {et~al.}(1981)\citenamefont
  {Alexander}, \citenamefont {Bernasconi}, \citenamefont {Schneider},\ and\
  \citenamefont {Orbach}}]{alexander1981excitation}%
  \BibitemOpen
  \bibfield  {author} {\bibinfo {author} {\bibfnamefont {S.}~\bibnamefont
  {Alexander}}, \bibinfo {author} {\bibfnamefont {J.}~\bibnamefont
  {Bernasconi}}, \bibinfo {author} {\bibfnamefont {W.~R.}\ \bibnamefont
  {Schneider}}, \ and\ \bibinfo {author} {\bibfnamefont {R.}~\bibnamefont
  {Orbach}},\ }\href {\doibase 10.1103/RevModPhys.53.175} {\bibfield  {journal}
  {\bibinfo  {journal} {Rev. Mod. Phys.}\ }\textbf {\bibinfo {volume} {53}},\
  \bibinfo {pages} {175} (\bibinfo {year} {1981})}\BibitemShut {NoStop}%
\bibitem [{\citenamefont {Ziman}(1982)}]{ziman}%
  \BibitemOpen
  \bibfield  {author} {\bibinfo {author} {\bibfnamefont {T.~A.~L.}\
  \bibnamefont {Ziman}},\ }\href {\doibase 10.1103/PhysRevLett.49.337}
  {\bibfield  {journal} {\bibinfo  {journal} {Phys. Rev. Lett.}\ }\textbf
  {\bibinfo {volume} {49}},\ \bibinfo {pages} {337} (\bibinfo {year}
  {1982})}\BibitemShut {NoStop}%
\bibitem [{\citenamefont {Amir}\ \emph {et~al.}(2010)\citenamefont {Amir},
  \citenamefont {Oreg},\ and\ \citenamefont {Imry}}]{Exp_Mat}%
  \BibitemOpen
  \bibfield  {author} {\bibinfo {author} {\bibfnamefont {A.}~\bibnamefont
  {Amir}}, \bibinfo {author} {\bibfnamefont {Y.}~\bibnamefont {Oreg}}, \ and\
  \bibinfo {author} {\bibfnamefont {Y.}~\bibnamefont {Imry}},\ }\href {\doibase
  10.1103/PhysRevLett.105.070601} {\bibfield  {journal} {\bibinfo  {journal}
  {Phys. Rev. Lett.}\ }\textbf {\bibinfo {volume} {105}},\ \bibinfo {pages}
  {070601} (\bibinfo {year} {2010})}\BibitemShut {NoStop}%
\bibitem [{zim()}]{ziman2}%
  \BibitemOpen
  \href@noop {} {}\bibinfo {note} {Note that in the notation of this work,
  $\gamma=1-\epsilon$ and $s=\omega^2$.}\BibitemShut {Stop}%
\bibitem [{\citenamefont {O'Connor}\ and\ \citenamefont
  {Lebowitz}(1974)}]{connor}%
  \BibitemOpen
  \bibfield  {author} {\bibinfo {author} {\bibfnamefont {A.}~\bibnamefont
  {O'Connor}}\ and\ \bibinfo {author} {\bibfnamefont {J.}~\bibnamefont
  {Lebowitz}},\ }\href@noop {} {\bibfield  {journal} {\bibinfo  {journal}
  {Journal of Mathematical Physics}\ }\textbf {\bibinfo {volume} {15}},\
  \bibinfo {pages} {692} (\bibinfo {year} {1974})}\BibitemShut {NoStop}%
\bibitem [{\citenamefont {Baluni}\ and\ \citenamefont
  {Willemsen}(1985)}]{baluni}%
  \BibitemOpen
  \bibfield  {author} {\bibinfo {author} {\bibfnamefont {V.}~\bibnamefont
  {Baluni}}\ and\ \bibinfo {author} {\bibfnamefont {J.}~\bibnamefont
  {Willemsen}},\ }\href {\doibase 10.1103/PhysRevA.31.3358} {\bibfield
  {journal} {\bibinfo  {journal} {Phys. Rev. A}\ }\textbf {\bibinfo {volume}
  {31}},\ \bibinfo {pages} {3358} (\bibinfo {year} {1985})}\BibitemShut
  {NoStop}%
\bibitem [{\citenamefont {Thouless}(1973)}]{thouless1973localization}%
  \BibitemOpen
  \bibfield  {author} {\bibinfo {author} {\bibfnamefont {D.}~\bibnamefont
  {Thouless}},\ }\href@noop {} {\bibfield  {journal} {\bibinfo  {journal}
  {Journal of Physics C: Solid State Physics}\ }\textbf {\bibinfo {volume}
  {6}},\ \bibinfo {pages} {L49} (\bibinfo {year} {1973})}\BibitemShut {NoStop}%
\bibitem [{\citenamefont {Klafter}\ and\ \citenamefont
  {Sokolov}(2011)}]{klafter2011first}%
  \BibitemOpen
  \bibfield  {author} {\bibinfo {author} {\bibfnamefont {J.}~\bibnamefont
  {Klafter}}\ and\ \bibinfo {author} {\bibfnamefont {I.~M.}\ \bibnamefont
  {Sokolov}},\ }\href@noop {} {\emph {\bibinfo {title} {First steps in random
  walks: from tools to applications}}}\ (\bibinfo  {publisher} {Oxford
  University Press},\ \bibinfo {year} {2011})\BibitemShut {NoStop}%
\bibitem [{\citenamefont {Bouchaud}(1992)}]{bouchaud1992weak}%
  \BibitemOpen
  \bibfield  {author} {\bibinfo {author} {\bibfnamefont {J.-P.}\ \bibnamefont
  {Bouchaud}},\ }\href@noop {} {\bibfield  {journal} {\bibinfo  {journal}
  {Journal de Physique I}\ }\textbf {\bibinfo {volume} {2}},\ \bibinfo {pages}
  {1705} (\bibinfo {year} {1992})}\BibitemShut {NoStop}%
\end{thebibliography}
%


\pagebreak

\end{document}